\documentclass[conference]{IEEEtran}
\IEEEoverridecommandlockouts
\usepackage{amsmath,amsfonts}
\usepackage{graphicx}
\usepackage{textcomp}
\usepackage{xcolor}

\usepackage{algorithm}
\usepackage{algpseudocode}
\usepackage{float}
\usepackage{wrapfig}
\usepackage{multicol}
\usepackage{tabularray}
\usepackage{multirow}
\usepackage{tabu}
\usepackage[inkscapelatex=false]{svg}
\usepackage{subfig}
\usepackage{enumitem}
\usepackage{pifont}
\usepackage{dblfloatfix}
\usepackage[hang]{footmisc}
\usepackage{lipsum}
\usepackage[utf8]{inputenc}
\usepackage[T1]{fontenc}
\usepackage{hyperref}

\setlist[itemize]{leftmargin=*,noitemsep, topsep=0pt}

\usepackage[compact]{titlesec}  
\titlespacing*{\section}{0pt}{0.15\baselineskip}{0.15\baselineskip}
\titlespacing*{\subsection}{0pt}{0.1\baselineskip}{0.1\baselineskip}
\titlespacing{\subsubsection}{0pt}{0.2em}{0cm}
\titlespacing{\paragraph}{0pt}{0.2em}{0.1cm}

\usepackage{tikz}

\usepackage[labelfont=bf,font=small]{caption} 

\algnewcommand{\LineComment}[1]{\State \(\triangleright\) #1}
\algnewcommand\algorithmicinput{\textbf{Input:}}
\algnewcommand\algorithmicoutput{\textbf{Output:}}
\algnewcommand\algorithmicbreak{\textbf{break}}
\algnewcommand\algorithmiccontinue{\textbf{continue}}
\algnewcommand\Break{\algorithmicbreak{} }
\algnewcommand\Continue{\algorithmiccontinue{} }
\algnewcommand\Input{\item[\algorithmicinput]}
\algnewcommand\Output{\item[\algorithmicoutput]}

\makeatletter
\def\BState{\State\hskip-\ALG@thistlm}
\makeatother

\algdef{SE}[EVENT]{Event}{EndEvent}[1]{\textbf{Event} #1}{\algorithmicend\ \textbf{event}}
\algdef{SE}[FUNCTION]{Function}{EndFunction}[1]{\textbf{Function} #1}{\algorithmicend\ \textbf{function}}
\algtext*{EndEvent} 
\algtext*{EndFunction} 
\algtext*{EndIf} 
\algtext*{EndFor} 
\algtext*{EndWhile} 

\definecolor{darkGreen}{HTML}{38761d}

\def\BibTeX{{\rm B\kern-.05em{\sc i\kern-.025em b}\kern-.08em
    T\kern-.1667em\lower.7ex\hbox{E}\kern-.125emX}}

\usepackage{pgfgantt}
\usepackage{xcolor}
\definecolor{minesprimary}{RGB}{33,49,77}
\definecolor{minesaccenta}{RGB}{146,162,189}
\definecolor{minesaccentb}{RGB}{210,73,42}

\definecolor{primarycolor}{RGB}{33,49,77}   
\definecolor{accentcolor}{RGB}{146,162,189} 
\definecolor{lightcolor}{RGB}{146,162,189} 
\definecolor{angrycolor}{RGB}{210,73,42}    
\definecolor{Accentcolor}{RGB}{210,73,42}    
\definecolor{bgfillcolor}{RGB}{206,213,221} 

\definecolor{limeyellow}{RGB}{228,255,100}
\definecolor{customGreen}{HTML}{38761d}
\usepackage[colorinlistoftodos,prependcaption,textsize=footnotesize,color=limeyellow,linecolor=magenta]{todonotes}
\presetkeys{todonotes}{inline}{}

\usepackage[]{hyperref}

\newcommand{\ie}{\textit{i}.\textit{e}.,}
\newcommand{\eg}{\textit{e}.\textit{g}.,}

\newcommand{\sysname}[1]{\textit{MC$^3$}}

\usepackage[a4paper, total={184mm,239mm}]{geometry}
\def\BibTeX{{\rm B\kern-.05em{\sc i\kern-.025em b}\kern-.08em
    T\kern-.1667em\lower.7ex\hbox{E}\kern-.125emX}}

\begin{document}

\title{\sysname{}: Memory Contention-based Covert Channel Communication on Shared DRAM System-on-Chips}

\author{\IEEEauthorblockN{Ismet Dagli}
\IEEEauthorblockA{\textit{Department of Computer Science} \\
\textit{Colorado School of Mines}\\
Golden, CO, USA \\
ismetdagli@mines.edu}\\
\IEEEauthorblockN{Yuanchao Xu}
\IEEEauthorblockA{\textit{Department of Computer Science} \\
\textit{University of California Santa Cruz}\\
Santa Cruz, CA, USA \\
yxu314@ucsc.edu}
\and
\IEEEauthorblockN{James Crea}
\IEEEauthorblockA{\textit{Department of Computer Science} \\
\textit{Colorado School of Mines}\\
Golden, USA \\
jcrea@mines.edu}\\
\IEEEauthorblockN{Sel{\c{c}}uk K{ö}se}
\IEEEauthorblockA{\textit{Department of ECE } \\
\textit{University of Rochester}\\
Rochester, NY, USA \\
selcuk.kose@rochester.edu}
\and
\IEEEauthorblockN{Soner Seckiner}
\IEEEauthorblockA{\textit{Department of ECE} \\
\textit{University of Rochester}\\
Rochester, NY, USA \\
sseckine@ur.rochester.edu}
\\
\IEEEauthorblockN{Mehmet E. Belviranli}
\IEEEauthorblockA{\textit{Department of Computer Science} \\
\textit{Colorado School of Mines}\\
Golden, CO, USA \\
belviranli@mines.edu}}
\maketitle

\begin{abstract}
Shared memory system-on-chips (SM-SoCs) are ubiquitously employed by a wide range of computing platforms, including edge/IoT devices, autonomous systems, and smartphones. In SM-SoCs, system-wide shared memory enables a convenient and cost-effective mechanism for making data accessible across dozens of processing units (PUs), such as CPU cores and domain-specific accelerators. Due to the diverse computational characteristics of the PUs they embed, SM-SoCs often do not employ a shared last-level cache (LLC). Although covert channel attacks have been widely studied in shared memory systems, high-throughput communication has previously been feasible only by relying on an LLC or by possessing privileged or physical access to the shared memory subsystem.\looseness=-1

In this study, we introduce a new memory-contention-based covert communication attack, \sysname{}, which specifically targets shared system memory in mobile SoCs. Unlike existing attacks, our approach achieves high-throughput communication
without the need for an LLC or elevated access to the system. 
We explore the effectiveness of our methodology by demonstrating the trade-off between the channel transmission rate and the robustness of the communication. 
We evaluate \sysname{} on NVIDIA Orin AGX, NX, and Nano platforms and achieve transmission rates up to 6.4 Kbps with less than 1\% error rate.\looseness=-1
\end{abstract}


\section{Introduction}

Mobile system-on-chips (SoCs) house multiple types of processing units (PUs), including general-purpose CPU cores and domain-specific accelerators (DSAs), such as GPUs and deep learning accelerators. With the proliferation of integrated DSAs, modern SoCs can provide cost-effective and energy-efficient execution, making them ideal candidates for in-the-field computing in many areas (mobile phones~\cite{xiong2021mobiledets}, smart home environments~\cite{jalal2019wrist} and autonomous systems~\cite{orin_bib}).\looseness=-1

An emerging architectural feature of modern SoCs (\eg{} NVIDIA's Orin~\cite{orin_bib}, Apple's M3~\cite{apple_m3}, Qualcomm's Snapdragon~\cite{snapdragon_888}) is a shared main memory where the data is stored for access by all PUs.  
The use of shared physical memory (SM) in commodity SoCs is motivated by the goal of reducing the chip area and production costs. 
It can also provide additional performance benefits by minimizing data transfer overhead between the CPU and the DSAs~\cite{dashti2017analyzing,dagli2022axonn}. However, several studies~\cite{xu2021pccs,hill2019gables,dagli2024shared} revealed that when running multiple workloads concurrently, PUs in SM-SoCs can experience significant slowdown caused by shared memory contention.\looseness=-1

Over the years, security researchers have shown that having a shared hardware component with a predictable performance slowdown leaves a unique fingerprint. Using this fingerprint, various attacks have exploited cache~\cite{liu2015last}, memory~\cite{zhenyu2012whispers}, storage~\cite{jiang2023sync+}, temperature~\cite{gonzalez2023first}, and power~\cite{taneja2023hot}. Covert channel communication attacks targeting vulnerabilities in the memory subsystem can be categorized into three: (1)~\textit{Cache-based, high-throughput attacks} which leverage the LLC between CPU cores~\cite{liu2015last,zhang2017attacks,gangwar2024mathcal,lipp2016armageddon,chen2024prime+}, between multiple GPUs~\cite{dutta2023spy} and between a CPU and a GPU~\cite{dutta2021leaky}. 
(2)~\textit{Low-throughput attacks targeting directly the DRAM} rely on memory performance attacks~\cite{schwarz2017fantastic}, memory deduplication~\cite{bosman2016dedup,xiao2013security}, bus snooping~\cite{zhenyu2012whispers} and monitoring of DRAM power consumption~\cite{paiva2019robust}.
(3) \textit{Attacks requiring elevated privileges or hardware access}~\cite{pessl2016drama,luo2016whispers,schwarz2017fantastic,schwarz2019zombieload}. None of these studies have shown how to build a fast, memory-contention-based covert channel without the need for privileged access (See Section~\ref{sec:thread_model}).\looseness=-1

Constructing a fine-grained, low-noise, and high-throughput memory-contention-based covert channel attack on mobile SM-SoCs presents several challenges: (i)~SM-SoCs generally lack a shared LLC to avoid complex design requirements. The trojan (\ie{} transmitter) needs to generate sufficient memory pressure that is observable by the spy (\ie{} receiver). CPU-based workloads in resource-limited SM-SoCs often fail to fully utilize the memory bandwidth even when all cores are used~\cite{xu2021pccs}. Therefore,  accelerators with higher memory demands, such as GPUs, should be employed. Meanwhile, the generated memory pressure should be low enough to minimize the risk of being detected by system defenses. (ii)~The total memory pressure exerted cumulatively by the spy and the trojan must be reliably high. Memory accesses satisfied by the caches can artificially increase perceived bandwidth, hence, they should be minimized to achieve a reliable access stream reaching the DRAM. This is also crucial to maximize the capacity of the communication channel. (iii)~Without external synchronization mechanisms, reliable and high-throughput data transmission over SM becomes challenging, as the trojan and the spy may operate at different magnitudes of memory operations, particularly when located on different types of PUs (\eg{} CPU cores and GPU). (iv) Finally, these requirements should be achieved without elevated privileges and hardware access, and the attack should function under single-user and multi-application environments. Our proposed work addresses all four challenges listed above.\looseness=-1

\begin{figure}[t!]
    \centering
    \includegraphics[width=0.91\linewidth]{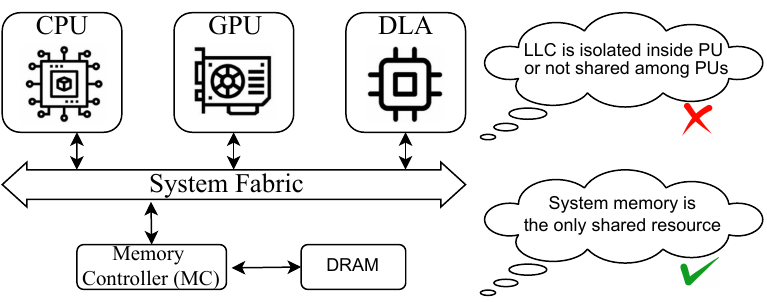}\\
    \vspace{-0.5em}
    \caption{Block diagram for NVIDIA’s Xavier AGX SoC embedding a CPU, GPU, deep learning accelerator (DLA) and shared memory.}
    \vspace{-1.7em}
    \label{fig:simplified_xavier_diagram}
\end{figure}

In this paper, we introduce a new  \underline{m}emory \underline{c}ontention-based \underline{c}overt \underline{c}ommunication attack, \sysname{}, targeting shared memory SoCs on mobile platforms. Our attack exploits the underlying vulnerability with software-only mechanisms, requiring neither direct hardware access nor super-user privileges. \sysname{} is designed to achieve a balance between the communication accuracy and the transmission rate (\ie{} capacity) of the covert channel. To increase the transmission rate and improve the efficiency of our attack, we further propose a CPU+GPU version of the receiver and the transmitter. We demonstrate that \sysname{} achieves transmission rates of up to 6.4 Kbps with an error rate below 1\% in CPU-to-GPU communication. Our implementation is available at \url{https://github.com/hypesys/MC3}.\looseness=-1

Our work makes the following contributions:
\begin{itemize}
    \item We unveil a new attack vector that leverages the slowdown in memory accesses due to shared use of system memory. The attack vector is achieved through software-only measurements and does not require privileged access to the system. 
    \item We present a novel covert channel attack that targets shared-memory SoCs without a last-level cache between its processing units. Our attack considers both CPU-GPU and CPU-CPU placements of the transmitter and the receiver.
    \item We evaluate \sysname{} on NVIDIA Orin AGX, Nano, and NX SoCs and achieve a channel capacity of up to 6.4 Kbps with 95\% accuracy. The accuracy reaches 99.99\% when the capacity is capped at 1.3 Kpbs.
    
\end{itemize}

\section{Background}

\subsection{Shared Memory SoCs (SM-SoC)}
\label{sec:smsoc}


Modern SoCs, such as NVIDIA’s Xavier and Orin architectures (as depicted in Fig.~\ref{fig:simplified_xavier_diagram}) integrate different types of accelerators, such as GPUs and DSAs, and each is optimized for specific computations. Unlike larger-scale systems where each accelerator has a dedicated primary memory, SM-SoCs share a common DRAM-based memory such as DDR4. Each PU has access to memory via a shared memory bus and a centralized memory controller (MC). Due to their inherent architectural heterogeneity, SM-SoCs often lack a shared LLC. Modern SoCs benefit from shared memory design because it reduces production costs and improves the data transfer overhead between CPU, GPU and other PUs.

\subsection{Additional Related Work}
Denial of service (DoS) attacks~\cite{mutlu2007memory, zhang2017attacks} exhaust the memory subsystem and greatly increase memory access latency. Commonly implemented memory controller (MC) scheduling policies, such as fairness control~\cite{ebrahimi2010fairness} and adaptive scheduler~\cite{kim2010atlas}, are typically designed to maximize system performance.
Although MC schedulers that prioritize security ~\cite{shafiee2015avoiding,wang2014timing} mitigate memory performance attacks with limited overhead, they cannot fully eliminate the threat. Our work, however, focuses on covert-channel communication, which demands far higher precision than DoS attacks require.

Additional studies have shown how to create architectural covert channels in the cloud~\cite{zhenyu2012whispers,zhang2017attacks}, HPC servers~\cite{dutta2023spy}, and desktops~\cite{matyunin2018tracking,green2017autolock}. Similar studies also used temperature~\cite{taneja2023hot,gonzalez2023first} and power~\cite{khatamifard2019powert,taneja2023hot} as a covert communication channel. Our work focuses on the vulnerabilities stemming from shared memory use.\looseness=-1

\section{Threat Model}
\label{sec:thread_model}

\begin{figure}[t]
    \centering
    \includegraphics[width=0.9\linewidth]{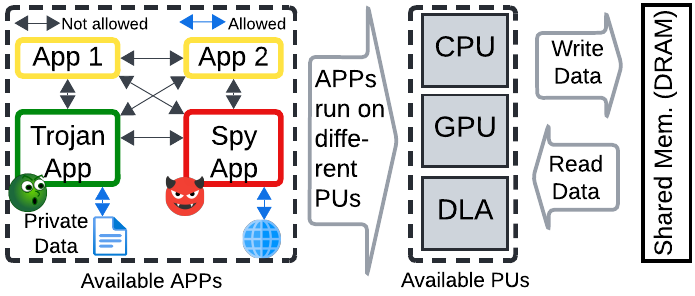}
    \vspace{-0.5em}
    \caption{Threat Model}
    \vspace{-1.8em}
    \label{fig:threat_model}
\end{figure}

Figure~\ref{fig:threat_model} depicts our threat model which involves running two (or more) applications on an SM-SoC (as explained in Section~\ref{sec:smsoc}). The transmitter (\ie{} trojan) is an application that has access to sensitive or private user data. The receiver (\ie{} spy) is an application running on the same SM-SoC but does not have access to the same data. Applications running in the system (including the receiver and transmitter) are not allowed to communicate with each other. Both transmitter and receiver can run on the CPU or GPU (in no specific order) without elevated execution privileges ---meaning they cannot access protected OS facilities or performance counters. The attack is designed to be executed remotely and attacker's presence or active engagement is not required.
The attacker is assumed to have no physical access to the hardware components (\eg{} for measuring power consumption and electromagnetic emissions).

\section{Shared Memory Contention as Attack Vector}

Our proposed methodology relies on the vulnerability that we discover in shared-memory SoCs and has the potential to be exploited on mobile and autonomous SoCs for a variety of attacks that do not require privileged access. A programmer can develop an adversarial transmitter application that leaves a distinct signature via purposeful shared memory accesses. This signature can be used to leak crucial information that can be encoded in binary form.






Although this attack strategy seems similar to other types of covert channel attacks, there are unique challenges to efficiently and reliably designing shared memory contention channels:\looseness=-1
\begin{itemize}
    \item \textit{Sufficiently observable contention:} While fully stressing memory resources to maximize contention is technically possible by the use of accelerators, the attack should generate sufficiently enough contention for the transmitter and receiver to communicate with each other. This will allow the attack to remain undetected by countermeasures deployed by the OS while maximizing the channel capacity of the attack. 
    We deal with this challenge by creating contention that only targets shared memory resources and bypasses the private cache hierarchy of CPUs.\looseness=-1
    \item \textit{Reliable and efficient contention:} Achieving reliable contention generation necessitates a careful characterization of the channel's behavior. While reliability can be increased by repeatedly performing contention for a long time to transmit a bit, the practicality of the attack often requires minimizing the repetition. We overcome these challenges by thoroughly analyzing the contention behavior 
    and using fine-grained time intervals for the receiver and the transmitter.
    \item \textit{Synchronizing transmitter and receiver:} Unlike traditional cache attacks, where the effects of a cache hit-or-miss can be clearly observed in the order of nanoseconds, the slowdown caused by the memory contention becomes visible in the order of microseconds. This requires synchronized transmitter and receiver operation. Additionally, considering the differences in computational capabilities and clock rates of the CPUs and GPUs, the design of attack vectors on two diverse PUs requires the synchronization to be done without using any external resources. We overcome this challenge by developing a precise contention generator and sleep procedure for the transmitter and adapting them to the receiver accordingly.
\end{itemize}

\begin{figure}[t]
    \centering
    \includegraphics[width=\linewidth]{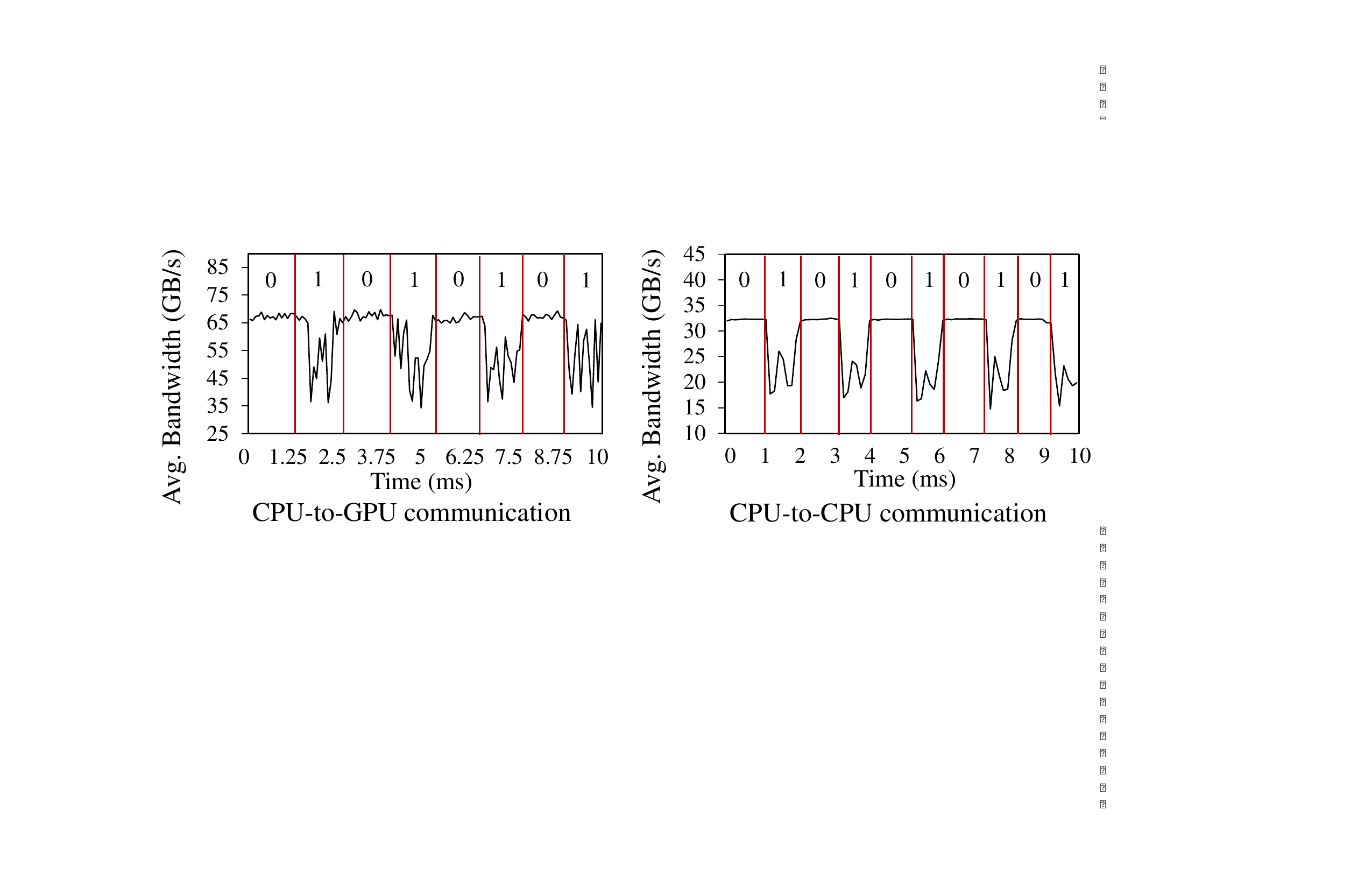}
    \vspace{-1.5em}
    \caption{Raw traces for CPU-to-GPU and CPU-to-CPU communication}
    \vspace{-2em}
    \label{fig:ViabilityExperiments}
\end{figure}


\noindent\textit{Feasibility of shared memory contention-based covert channel:} 
To demonstrate how the memory contention behavior affects the observed memory bandwidth of an application, we run the transmitter app on the CPU and the receiver on both the CPU and the GPU of an Orin NX SoC. Fig.~\ref{fig:ViabilityExperiments} shows two raw traces of the varying average bandwidth (BW) perceived on the receiver side. Regardless of whether the receiver runs on the CPU or the GPU, the perceived BWs for the receiver have clear drops in the traces, which correspond to the `1's sent by the transmitter. This experiment demonstrates the feasibility of building covert channels with shared memory contention.


\section{\sysname{}: Shared Memory Contention-based \\Covert Channel Communication}

\subsection{Overall Mechanism}

\begin{figure}[b]
    \vspace{-2em}
    \centering
    \includegraphics[width=0.9\linewidth]{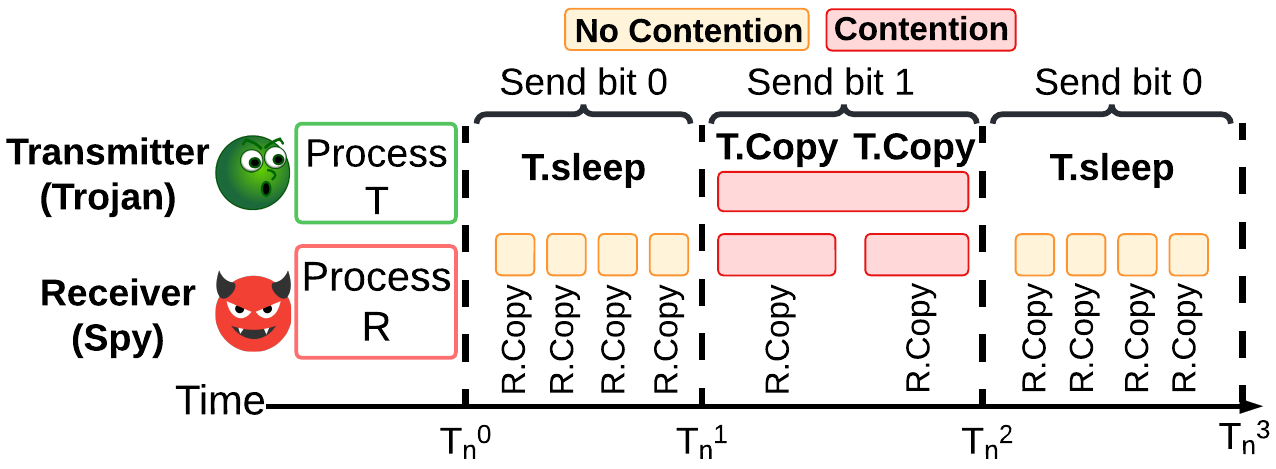}
    \vspace{-0.5em}
    \caption{Communication between the transmitter and receiver.}
    \label{fig:bit0and1_illustration}
\end{figure}

Fig.~\ref{fig:bit0and1_illustration} illustrates the communication protocol between the transmitter and receiver for transmitting bits (\ie{} 0 or 1) through shared memory contention. The transmitter conveys bits by performing buffer copy operations on memory while the receiver continuously performs another buffer copy operation to detect the transmitted bit.

The transmitter is responsible for sending the bit by modulating the level of memory contention. As shown in the upper part of Fig.~\ref{fig:bit0and1_illustration}, to transmit a bit `0', the transmitter sleeps for a predefined time interval of $T_{n}$. To send a bit `1', it performs continuous copy operations to access DRAM. The receiver continuously operates its own buffer copy function, then measures the duration of the buffer copy operation, and finally calculates the average BW over the duration, which will be used to decode the bit. 

The lower part of Fig.~\ref{fig:bit0and1_illustration} illustrates the case where the receiver can perform more copy operations with lower latency (\ie{} higher memory BW) while the transmitter is sleeping. Multiple copy operations per time interval $T_{n}$ can also be utilized to have more reliable data transmission (see Sec.~\ref{sec:reducing_noise}). In contrast, when the transmitter induces contention by performing a copy operation, the receiver's throughput decreases (\ie{} lower memory BW) because memory latency increases due to shared memory contention.\looseness=-1

While it is possible to run the receiver non-stop, we instead opt to start the receiver slightly earlier (\eg{} one second) than the transmitter at a predetermined epoch. This design decision eliminates the need for continuous operation of the receiver, thus minimizing the chances of our attack being detected under real-world conditions. The early start allows the receiver to collect BW information (\ie{} latency per copy operation) without any interference from the transmitter, which can be used as a baseline during the data analysis stage. Although other applications on the device may use shared memory ---potentially introducing noise into the receiver's average BW measurements--- the pattern of zeros and ones can be detected using heuristic history-based signal processing approaches.

Table~\ref{tab:devices} lists our test devices with varying computational capability and memory BW capacity. We use Jetpack 5.1 on all devices. It is worth noting that all devices have TrustZone Trusted Execution Environment (TEE) and OS-protection regions in the memory subsystem.

\begin{table}[t]
\centering
\caption{Targeted platforms}
\vspace{-0.5em}
\label{tab:devices}
\begin{tabular}
{|m{2.35em}|m{7.45em}|m{7.9em}|m{7.9em}|}
\hline
\textbf{Device} & \multicolumn{1}{c|}{Orin AGX} & \multicolumn{1}{c|}{Orin Nano} & \multicolumn{1}{c|}{Orin NX} \\ \hline
CPU & 12-core A78 & 6-core A78AE & 8-core A78AE \\ \hline
GPU & 2048 core Ampere & 1024 core Ampere & 1024 core Ampere \\ \hline
\begin{tabular}[c]{@{}l@{}}DRAM \\ \& BW\end{tabular} & \begin{tabular}[c]{@{}l@{}}64 GB 256-bit\\ 204.8 GB/s\end{tabular} & \begin{tabular}[c]{@{}l@{}}8 GB 128-bit\\ 68 GB/s\end{tabular} & \begin{tabular}[c]{@{}l@{}}8 GB 128-bit\\ 102 GB/s\end{tabular} \\ \hline
\end{tabular}
\vspace{-2.0em}
\end{table}

\subsection{Attack Vector}\label{sec:Attack_vector}
Our attack leverages a memory-contention channel to stealthily transmit data between two processes.

Our transmitter algorithm, described in Alg.~\ref{alg:transmitter}, is responsible for encoding the data and transmitting the encoded data bitstream over the memory-contention channel. It loops through the input data bitstream, either running the memory contention kernel (\ie{} data copy operation) for the specified duration $T$ if the current bit is a $1$, resulting in memory contention, or sleeping for the same duration $T$ (if not set differently) if the current bit is a $0$, resulting in near zero memory contention.\looseness=-1

Our receiver algorithm, described in Alg.~\ref{alg:receiver}, is responsible for receiving the encoded data from the memory-contention channel and decoding it. It continuously runs the memory contention kernel over the duration $T$ ---the $T$ value should be the same as the transmitter's. Then, it normalizes the BWs based on a global average (\ie{} subtract each BW sample from the overall average observed BW)~\cite{zhenyu2012whispers}. Then, this normalized BW is thresholded with hysteresis, which we experimentally determined to be much more resistant to noise. Finally, the result of this thresholding is converted directly into the received bitstream, where the values above the threshold become $0$ (which corresponds to a higher measured BW in the receiver, as a result of the transmitter \textit{not} simultaneously generating memory contention) and the values below the additive inverse (\ie{}  $-1 \times$) of the threshold are $1$ (which corresponds to a lower measured BW in the receiver, as a result of the transmitter simultaneously generating memory contention).\looseness=-1

\vspace{-0.5em}

\begin{algorithm}[h]
\caption{Transmitter}\label{alg:transmitter}
\small
\begin{algorithmic}
\Input{data bitstream $B$ and its length $n$, time interval $T$}
\For{$i \gets 0$ to $n - 1$}
    \If{$B[i]$ is $1$}
        \Call{contend\_for}{$T$} \Comment{Generate contention}
    \Else
        \;\Call{sleep\_for}{$T$} \Comment{Remain idle}
    \EndIf
\EndFor
\end{algorithmic}
\end{algorithm}

\vspace{-1.5em}

\begin{algorithm}[h]
\caption{Receiver}\label{alg:receiver}
\small
\begin{algorithmic}
\Input{hysteresis threshold $\gamma$, run length $n$, time interval $T$}
\State $B \gets []$ \Comment{Output bitstream (starts empty)}
\State $b \gets 0$ \Comment{Hysteresis state}
\State $\bar{\beta} \gets 0$ \Comment{Average BW}
\For{$i \gets 0$ to $n - 1$}
    \State $\beta_{\text{raw}} \gets$ \Call{contend\_for}{$T$}
    \State $\bar{\beta} = \frac{\bar{\beta} \cdot i + \beta_{\text{raw}}}{i + 1}$ \Comment{Use simple global average}
    \State $\beta_{\text{normalized}} = \beta_{\text{raw}} - \bar{\beta}$
    \If{$\beta_{\text{normalized}} > \gamma$}
        \Call{append}{$B$, 0}
        \State $b \gets 1$
    \ElsIf{$\beta_{\text{normalized}} < (-1 \cdot \gamma)$}
        \Call{append}{$B$, 1}
        \State $b \gets 0$
    \Else
        \;\Call{append}{$B$, $b$}
    \EndIf
\EndFor
\Return $B$
\end{algorithmic}
\end{algorithm}
 \vspace{-0.5em}

\subsection{Cache-less Memory Access}


Throughout the development of the memory-contention kernel, we observed that CPU caches are used to access the data and artificially inflate the memory BW measurements by using data-streaming kernels. Although sufficient contention generation using data streaming kernels is also possible, it makes our BW measurements unreliable and introduces substantial noise into the communication channel.

To alleviate this, our implementation employs memory instructions with non-temporal hints, specifically \verb|ldnp|/\verb|stnp| (Load/Store Pair Non-Temporal) Arm64 instructions. This hint signifies that the data being loaded or stored is unlikely to be reused soon, prompting the system to bypass the cache hierarchy~\cite{arm2024manual}. By doing so, we ensure that our memory operations access DRAM directly, enhancing the reliability of our BW measurements and increasing the efficiency of the attack. It is worth noting that we also observed sufficient contention generation with data streaming using regular data streaming without non-temporal instructions.






\begin{table}[t]
\caption{Precise `contention'  and `sleep for' durations}
\label{tab:precise_contention_sleep}
\vspace{-0.5em}
\begin{tabular}{|l|l|l|l|l|l|}
\hline
Operation & \begin{tabular}[c]{@{}l@{}}Expected \\ duration\end{tabular} & \begin{tabular}[c]{@{}l@{}}Mean \\ Error\end{tabular} & \begin{tabular}[c]{@{}l@{}}Minimum \\ Error\end{tabular} & \begin{tabular}[c]{@{}l@{}}Maximum \\ Error\end{tabular} & \begin{tabular}[c]{@{}l@{}}Std.\;dev. \\ Error\end{tabular} \\ \hline
Sleep for & $100\;\text{ms}$ & $46\;\text{ns}$ & $5\;\text{ns}$ & $314\;\text{ns}$ & $41\;\text{ns}$ \\ \hline
Contention & $100\;\text{ms}$ & $12\;\mu\text{s}$ & $5\;\text{ns}$ & $767\;\mu\text{s}$ & $65\;\mu\text{s}$ \\ \hline
\end{tabular}
\vspace{-2em}
\end{table}

\begin{figure*}[h!]
    \centering
    \vspace{-1.3em}
    \includegraphics[width=\linewidth]{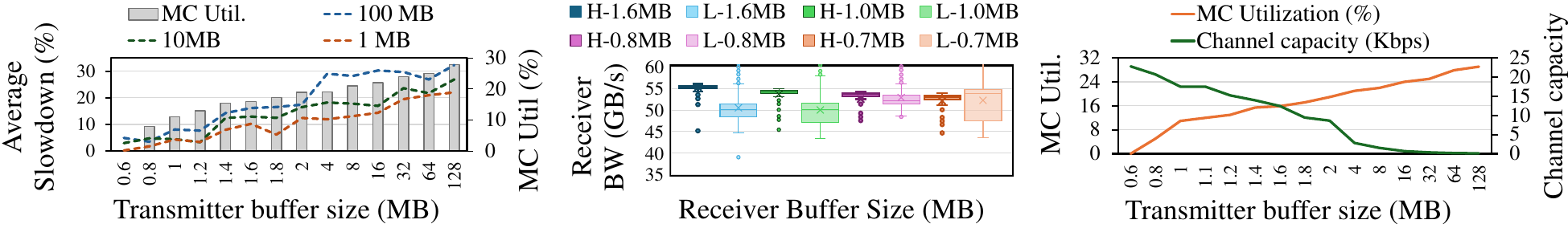}\\
    \vspace{-0.1em}
    \caption{(a) [Left] Average slowdown in the perceived BW depending on the transmitter buffer size. (b) [Middle] Average perceived high BW (H) and low BW (L) for bits `0' and `1', respectively, for varying receiver buffer sizes. (c) [Right] MC utilization per transmitter buffer size. }
    \vspace{-1.5em}
    \label{fig:buffer_size_experiments}
\end{figure*}

\subsection{Precise Contention Duration and Precise Sleep}

In order to maximize performance, the sleep and data copy operations require high temporal precision (\ie{}  sleep or run for the desired duration as accurately as possible). We implemented a precise sleep mechanism by utilizing an OS-provided function (\ie{}  \verb|std::this_thread::sleep_for|) until near the desired end time and finally spinning (\ie{}  \verb|while| loop with an empty body) until the desired end time is reached~\cite{blatnik2024sleep}. To implement \texttt{CONTEND\_FOR(T)}, used in both algorithms, our data copy operation first runs the memory-contention kernel for a small, fixed amount of data (\ie{} taking nearly one second) to estimate the currently achievable memory contention BW ($\beta_{0}$). Using the total desired duration ($T$), it estimates the amount of data the kernel needs to run for ($d_{*}$, where $d_{*} \propto \beta_{i} \cdot T$). Using this estimate, it splits this total data estimate up (e.g.: $d_{i} = \frac{1}{100} \cdot d_{*}$), runs the memory-contention kernel (for $d_{i}$ amount of data), collects $\beta_{i}$, updates $d_{*}$ and $d_{i}$, and repeat. When it gets close to the desired end time, it further splits the estimate (e.g.: $d_{i} = \frac{1}{1000} \cdot d_{*}$) to reduce the amount of under/overshoot. In Table~\ref{tab:precise_contention_sleep}, we report the results for 100 ms execution, demonstrating our average error rate of 4 and 7 orders of magnitude lower for sleep and contention generation, respectively.\looseness=-1

\subsection{Transmitter and Receiver Design}

Our attack hinges on the transmitter generating sufficient memory pressure to create noticeable contention, which the receiver must detect. Essentially, the transmitter needs to generate enough memory access requests to contend with the receiver, and the receiver must be sensitive enough to observe the difference between the transmitter's sleep and data copy operations. To evaluate this, we conducted experiments on an Orin Nano using three CPU cores for both the transmitter and receiver. We send 1024 bits of information, evenly distributed with bits `0's and `1's, with the slowdown results illustrated in Fig.~\ref{fig:buffer_size_experiments}.a. We design varying buffer sizes of data copy for transmitter (from 0.6 MB to 128 MB) and receiver (from 1 MB to 100 MB). Overall, we observe an increasing pattern of average slowdown on the receiver side as we increase transmitter buffer sizes. For example, with a 1 MB buffer size for the receiver, the transmitter requires at least a 2 MB buffer (\ie{} at least 20\% MC utilization) to achieve a minimum of 10\% slowdown.


Although we observe a slowdown while transmitting bit `0', we also need to clearly distinguish the BW differences between bit `1' and bit `0' on the receiver. To demonstrate this, we measure and analyze the average BW perceived on the receiver side while the transmitter is running, with a 4 MB buffer size sending `0' (higher perceived BW) and `1' (lower perceived BW) bits. Fig.~\ref{fig:buffer_size_experiments}.b depicts the distribution (including outliers) of the BW perceived by the receiver while sensing `0's (light colors) and `1's (dark colors). Although buffer sizes of 1.6 MB and 1.0 MB have clear differences in terms of perceived BW, lower buffer sizes for the receiver fail to distinguish the differences between bit `0' and `1' by looking at perceived BW. While outliers create noise if accuracy is calculated solely with average-based methods, history-based methods (which compare the current trace with the previous) clearly identify the changes.  Although the transmitter with 0.8 MB buffer has approximately 10\% MC utilization, contention may not be enough to distinguish bit `0's and `1's.\looseness=-1



\begin{figure}[b]
    \vspace{-1.5em}
    \centering
    \includegraphics[width=\linewidth]{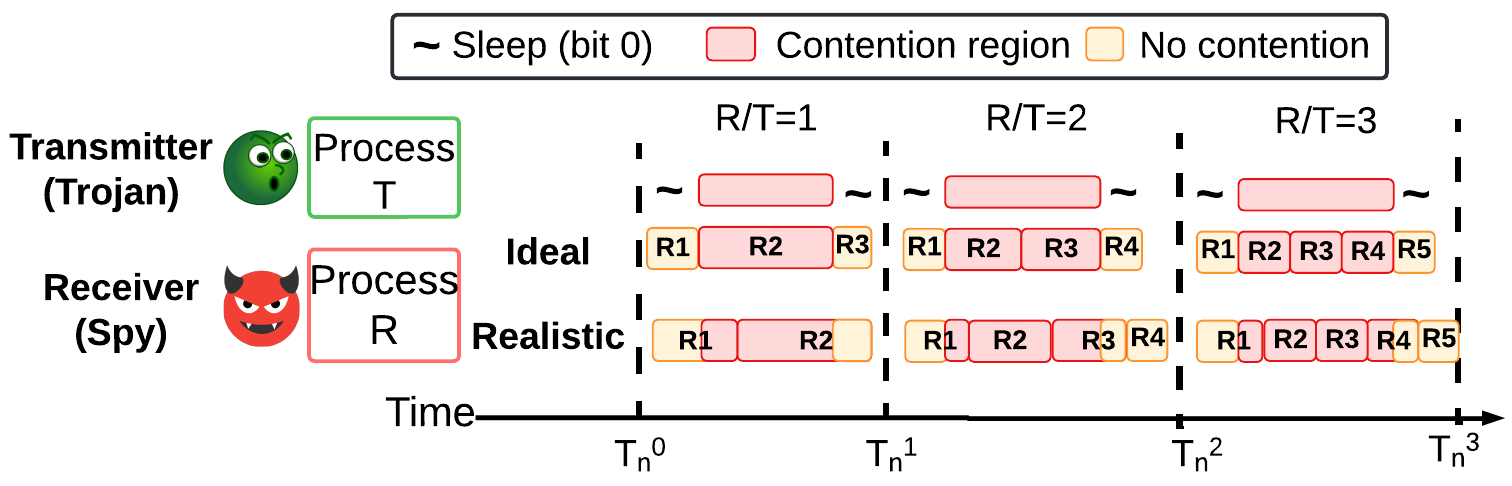}
    \vspace{-1.5em}
    \caption{The overlap between [T]ransmitter's and [R]eceiver's copy operations for various R/T copy epoch ratios. `Ideal' represents the expected durations and `actual' represents the observed. R1-R5 indicates the epoch number of a copy operation performed by the receiver. \looseness=-1
    }
    \label{fig:transmitter_receiver_time_intervals}
\end{figure}

\subsection{Trade-off between Copy Duration and Contention Amount}

The channel capacity intuitively depends on the size of the transmitter's buffer that is being copied. Assuming that, to transmit bit `1', transmitter copy operation with a fixed buffer size will be performed once during time interval $T_{n}$, there exists an inverse relationship between the transmitter buffer size and the channel capacity, as demonstrated in Eq.~\ref{eq:trade_off_buffer_time}. As we increase the transmitter buffer size, thus the time $Time_{Tra}$ to copy the buffer, and account for the average slowdown $Slowdown_{Rec}$ perceived by the receiver, the channel capacity decreases. 

\vspace{-1.8em}
\begin{equation}
    Channel \; Capacity = 1 \; / \; (Time_{Tra}*Slowdown_{Rec})
    \label{eq:trade_off_buffer_time}
\end{equation}
\vspace{-1.8em}

Ideally, we aim to use the smallest possible transmitter buffer size to maximize the channel capacity of our covert channel. On the other hand, there is a lower limit to how much we can decrease the buffer size to generate observable contention.  To demonstrate the trade-off between channel capacity and buffer size, we vary the transmitter buffer size and report the results in Fig.~\ref{fig:buffer_size_experiments}.c. We observe that increasing the transmitter buffer size leads to a near-linear decrease in channel capacity. For example, with transmitter buffer sizes of 0.5 MB and 0.6 MB, we achieved channel capacities of up to 25 Kbps, yet the receiver was unable to detect the transmitter's activity since the transmitter's MC utilization was nearly zero.


\subsection{Reducing Noise}\label{sec:reducing_noise}

Synchronizing transmitter and receiver is essential for accurately sensing BW differences. Since direct communication between the transmitter and receiver is not allowed (which would otherwise defeat the point of a covert channel), we must statically decide the time intervals. However, as depicted in Fig.~\ref{fig:buffer_size_experiments}.b, due to contention generation being inherently noisy, the accumulation of outlier-induced errors can lead to desynchronization. If we aim to operate in the worst-case scenario, then many contended regions may complete earlier than anticipated.

We illustrate and compare the expected (\ie{} ideal) and observed (\ie{} actual) copy operation durations in Fig.~\ref{fig:transmitter_receiver_time_intervals}. \textit{R/T} denotes the ratio of copy epochs (or iterations) that the (R)eceiver performs for every bit `1' sent by the (T)ransmitter. 
When \textit{R/T = 1}, the receiver perceives the contention from the transmitter with a delay, causing a drop in observed BW and increasing in noise 
Conversely, when the transmitter sends bit `1' across multiple epochs (\ie{} \textit{R/T > 1}), the receiver will capture at least one or more fully contended regions. This substantially improves transmission accuracy, while reducing the communication capacity of the covert channel.

\section{Improving channel capacity with GPUs}

Mobile and autonomous SoCs often embed GPUs which are designed to enable massive parallelism. This results in better utilization of the memory subsystem compared to the CPUs.

\begin{figure}[b]
    \vspace{-1.3em}
    \centering
    \includegraphics[width=\linewidth]{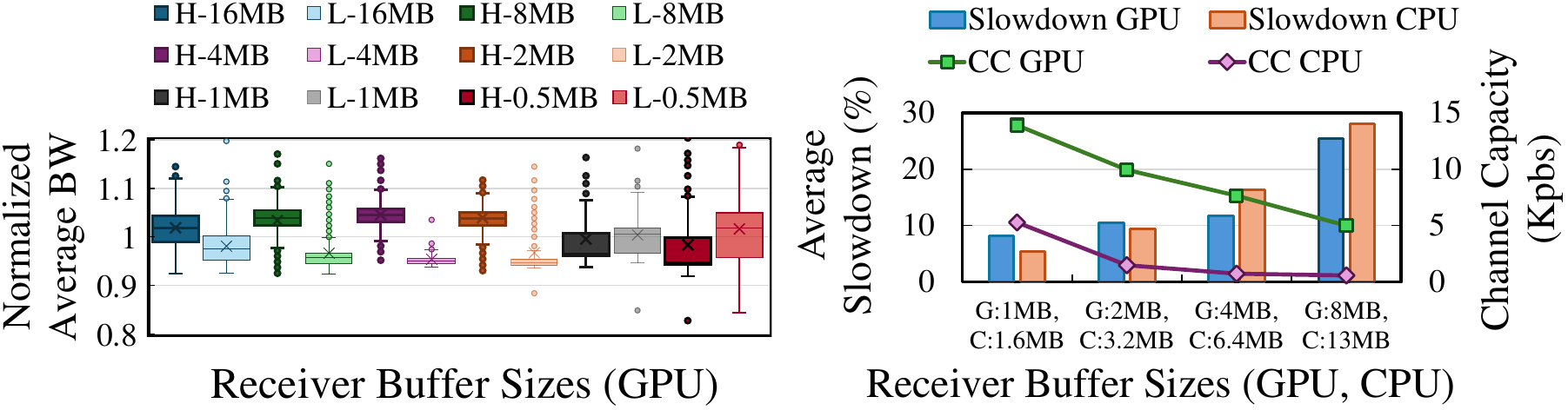}\\
    \caption{ (a)~[Left]~The distribution of the perceived BW distribution for varying buffer sizes. H and L indicates `0' and `1' transmissions, respectively.  (b)~[Right]~Slowdown in the perceived BW and channel capacity when receiver is on (C)PU and (G)PU.}
    \label{fig:merged_receiver_gpu_contention_sensing}
\end{figure}

\begin{figure*}[t]
    \centering
    \includegraphics[width=\linewidth]{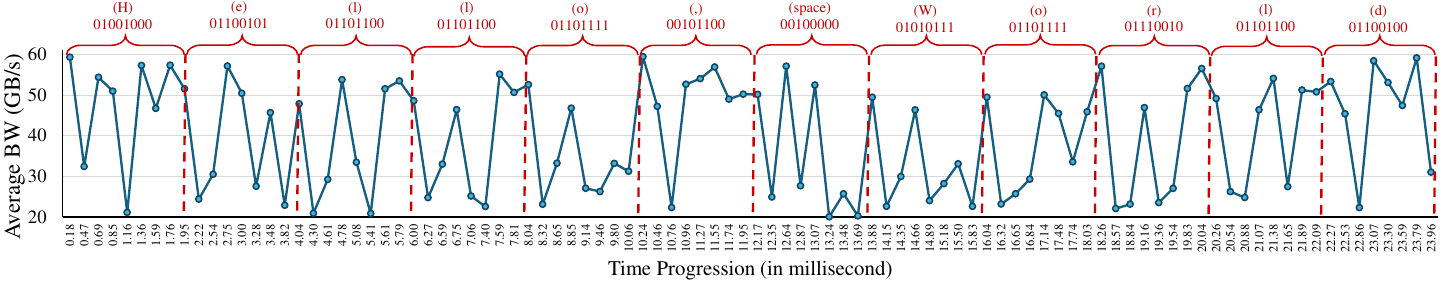}\\
    \vspace{-0.5em}
    \caption{Average BW observed by the receiver while receiving a "Hello, World" message.}
    \vspace{-1.5em}
    \label{fig:hello_world}
\end{figure*}
\subsection{Receiver on the GPU}

The receiver must generate sufficiently high BW to accurately distinguish between a `0' bit and a `1' bit. To achieve better accuracy, we run the receiver on the GPU and the transmitter on the CPU. We implement the memory copy operation with \texttt{cudaMemcpy} using CUDA~\cite{cuda_samples}. Fig.~\ref{fig:merged_receiver_gpu_contention_sensing}.a shows the average receiver slowdown for the `0' (\ie{} high) and `1' (\ie{} low) bits using a buffer size of 1 MB on an Orin AGX. For receiver buffer sizes from 2MB to 8MB, we observe a clear distinction between `0's and `1's. However, buffer sizes of 0.5MB and 1MB fail to sense the contention, whereas buffer sizes of 16MB and beyond may result in a misalignment of the copy epochs 
as explained in Sec.~\ref{sec:reducing_noise}. We also experimented with configuring the transmitter on the GPU and the receiver on the CPU. However, unlike the case where the receiver is on the GPU, placing the transmitter on the GPU did not improve the distinction of `1' and `0' bits. These results were omitted because of space limitations.

\subsection{Channel Capacities for Receivers on the GPU and CPU}
To further understand the relationship between channel capacity and the slowdown observed when the receiver is placed on CPU and GPU, we perform an experiment on Orin AGX
where we gradually increase receiver buffer sizes and
report the results in Fig.~\ref{fig:merged_receiver_gpu_contention_sensing}.b. We observe that we can achieve channel capacities of up to 14 Kbps and 5 Kbps on GPU and CPU, respectively, with 9\% and 7\% slowdowns observed in the measured BW. As we increase buffer sizes, we typically observe less channel capacity but more sensible contention on both CPU- and GPU-based receivers. It is worth noting that, when the receiver runs on GPU, we can map the transmitter to 11 cores out of the 12 available CPU cores. 
This mapping significantly increases the contention generation capacity of the transmitter. Overall, the GPU-based receiver achieves approximately 3× – 5× higher channel capacities on Orin AGX (and 2× – 4× on Orin Nano) compared to the CPU-based receiver.\looseness=-1


\begin{figure}[b]
    \vspace{-1em}
    \centering
    \includegraphics[width=\linewidth]{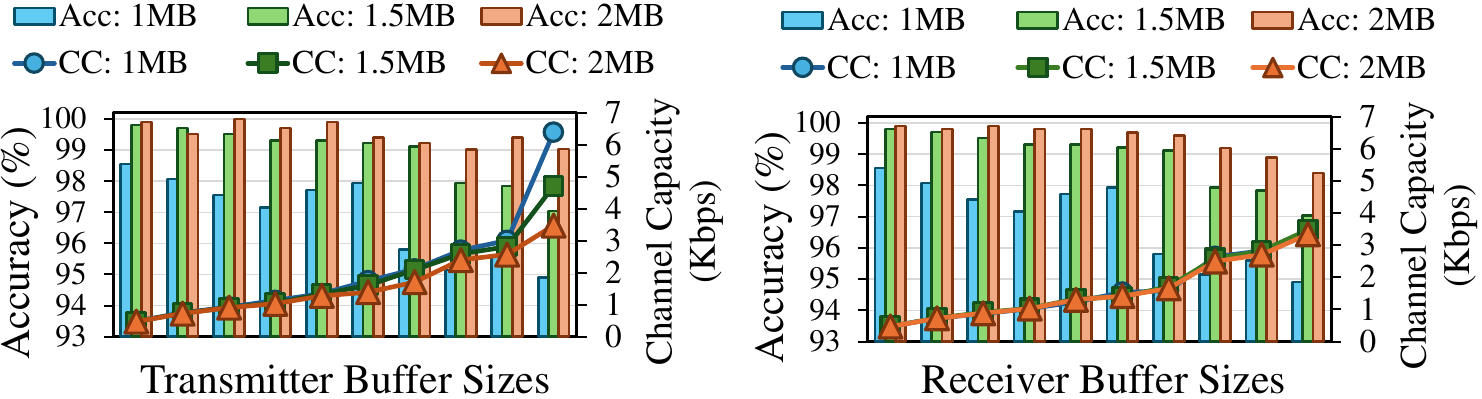}\\
    \caption{The trade-off between accuracy and channel capacity for varying (a) transmitter [left] and (b) receiver buffer sizes [right].}
    \label{fig:accuracy_transmitrate_gpu}
\end{figure}

\subsection{Hello World Transmission}

To assess our design's performance with longer messages, we transmit a 100 Kb text message on Orin Nano and observe how the perceived BW changes over time. The results are depicted in Fig.~\ref{fig:hello_world}. The y-axis shows the rolling average of perceived BW during each time interval, whereas the x-axis represents the time progression in milliseconds. The initial portion of the message is transmitted and received with 100\% accuracy while the entire message is delivered with 99.02\% accuracy at a channel capacity in excess of 4 Kbps. \looseness=-1



\subsection{Channel Capacity vs. Transmission Accuracy}
As the final overarching experiment, we vary the transmitter and buffer sizes and observe the resulting trade-off between channel capacity and transmission accuracy when the receiver is run on the GPU and the transmitter on the CPU of Orin AGX. 
In Fig.~\ref{fig:accuracy_transmitrate_gpu}.a, we varied the transmitter buffer sizes while keeping the receiver buffer size fixed at 1 MB. We increase the channel capacity by varying the number of copy epochs/iterations of the buffer copy operation per time interval from 1 to 10, and adjusting the receiver accordingly. In general, our results demonstrate that \sysname{} achieves either up to 6.4 Kbps channel capacity or up to 99.99\% transmission accuracy. As the transmitter buffer size and the number of copy iterations per interval increased, we observe higher accuracy but reduced channel capacity. Some notable data points are:
\begin{itemize}
\item The 2MB transmitter buffer size achieves 99. 1\% accuracy at 3.5 Kpbs channel capacity and a near-perfect accuracy of 99.99\% at 1.3 Kpbs.
\item The 1MB transmitter buffer size maximizes channel capacity up to 6.4 Kpbs while achieving a decent 94.9\% accuracy.
\end{itemize}

In Fig.~\ref{fig:accuracy_transmitrate_gpu}.b, we increase receiver's buffer size and buffer copy operation iterations per interval, but keep the transmitter buffer size constant at 1 MB. Overall, increasing the receiver buffer size (with a constant transmitter copy iteration) improved accuracy with minimal impact on channel the capacity. It is worth noting that increasing the receiver size degrades the accuracy since \textit{R/T} ratio becomes unbalanced once the buffer size is 5 MB and beyond. Similar to the observations in Fig.~\ref{fig:accuracy_transmitrate_gpu}.a, increasing the channel capacity by decreasing the transmitter copy operations per interval leads to a decrease in accuracy.\looseness=-1










\section{Conclusion}

In conclusion, we demonstrate a novel and efficient covert channel attack that exploits shared-memory contention in SM-SoCs. The proposed attack does not require privileged access to the system and achieves a channel bandwidth of up to 6.4 Kbps with accuracy rates that reach 99.99\%. We unveil an important vulnerability that could be used to leak private data in modern mobile and autonomous systems.

\section*{Acknowledgements}
This work is supported in part
by the National Science
Foundation (NSF) under Grant No. CNS-2350228. Any opinions,
findings, or recommendations expressed in this material are those
of the authors and do not necessarily reflect the views of NSF.\looseness=-1

\clearpage

\bibliographystyle{plain}
\bibliography{references}

\end{document}